# Pseudogap of Superconducting correlation origin in iron-pnictide superconductor $Ba_{0.6+\delta}K_{0.4-\delta}Fe_2As_2$

Yong Seung Kwon[1,†], Jong Beom Hong[1], Yu Ran Jang[1], Hyun Jin Oh[1], Yun Young Song[1], Byeong Hun Min[1], Takuya Iizuka[2], Shin-ichi Kimura[2,3], A. V. Balatsky[4] and Yunkyu Bang [5,†]

[1]*Department of Physics, Sungkyunwan University, Suwon 440-746, Republic of Korea*

[2]*School of Physical Sciences, the Graduate University for Advanced Studies (SOKENDAI), Okazaki 444-8585, Japan*

[3]*UVSOR Facility, Institute for Molecular Science, Okazaki 444-8585, Japan*

[4]*T-Division, Center for Integrated Nanotechnologies, Los Alamos National Laboratory, Los Alamos, New Mexico 87545, USA*

[5]*Department of Physics, Chonnam National University, Kwangju 500-757, Republic of Korea*

[†] E-mail: yskwon@skku.ac.kr (Y.S.K.); ykbang@jnu.ac.kr (Y.B.)

**Pseudogap (PG), a phenomenon of opening of gap like features above superconducting (SC) transition temperature ($T_c$), has been universally observed in the high-$T_c$ cuprates (HTC) [1,2], some heavy fermion superconductors [3], and iron pnictides [4-7]. Here, we report the observation of PG behavior in optical conductivity in an underdoped $Ba_{0.6+\delta}K_{0.4-\delta}Fe_2As_2$ ($T_c^{onset}$ =36 K) single crystal far above $T_c$, up to 100 K (~ $3T_c$). Unique scale separation in $Ba_{0.6+\delta}K_{0.4-\delta}Fe_2As_2$ compound – magnetic and SC correlation energy scales being widely separated – enabled us to establish that the PG structures observed in the range of 50 $cm^{-1}$ - 150 $cm^{-1}$ in optical conductivity is caused by the SC correlation and the magnetic correlation origin is ruled out.**

**Theoretical calculations, based on the preformed Cooper pair model [8], provided an excellent description to the temperature evolution of the optical conductivity data from below to above $T_c$.**

We measured the optical spectroscopy in the slightly underdoped $Ba_{0.6+\delta}K_{0.4-\delta}Fe_2As_2$ ($T_c^{onset}$ =36 K) for a broad range of temperatures from 4 K to 300 K and frequencies from 15 cm$^{-1}$ to 28,000 cm$^{-1}$. The key advance of our spectroscopic measurement is the achievement of the high resolution terahertz (THz) spectroscopy down to 15 cm$^{-1}$ and the combined IR spectroscopy to cover the wide frequency range. With this technique the common limitation of the IR spectroscopy – which cannot provide reliable data below ∼ 100 cm$^{-1}$ that is often the most interesting frequency regime for the study of the SC correlation – has been overcome. We used Michelson-type rapid-scan Fourier spectrometer (JASCO FTIR610) in the frequency range of 40 cm$^{-1}$ - 12000 cm$^{-1}$ and the Martin-Puplett-type rapid scan Fourier spectrometer (JASCO FARIS) for the THz frequency range of 15 cm$^{-1}$ - 200 cm$^{-1}$ for temperatures of 4 K - 300 K. Overall uncertainties of the data in frequency ranges below and above 100 cm$^{-1}$ are maintained less than 0.6 % and 0.3 %, respectively. For technical details of the reference setting and control of data uncertainty, we refer to the supporting online material. High quality single crystals of $Ba_{0.6+\delta}K_{0.4-\delta}Fe_2As_2$ were grown by the Bridgeman method with a sealed tungsten crucible [9] at 1550 °C. The ratio of room temperature to residual resistivity (RRR), $\rho(300\ K) / \rho(T_c)$, is 7.5 and the sharp SC transition is observed at $T_c^{onset}$ = 36 K in the dc-resistivity data in Fig.1. Compared to $T_c$ = 38 K of the optimal doped $Ba_{0.6}K_{0.4}Fe_2As_2$ [10], our sample is a slightly underdoped one. No evidence of the SDW transition, such as discontinuity in the resistivity and magnetic susceptibility, is observed in our sample.

Figure 2 shows the raw data of $R(\omega, T)$ for different temperatures from 4 K to 300 K. It clearly shows new features below 150 cm$^{-1}$, the region that was not accessible to the previous IR spectroscopy studies [10, 11]. The reflectivity data below $T_c$ of 4, 8, 20, and 30 K become flat approaching $R \rightarrow 1$ at low frequencies reflecting the presence of a SC gap and show systematic drop starting from $\omega \approx 50$ cm$^{-1}$ indicating the beginning of absorptions. This low frequency feature in 50 cm$^{-1}$- 150 cm$^{-1}$ is beyond the experimental uncertainty mentioned above and the quality of our data should be compared to the data of Refs.*(10, 11)*. This liability and control of the uncertainty level of the data at low frequencies are of ultimate importance for studying the SC and PG features [12]. The broad depletion in the region of 200 cm$^{-1}$ - 800 cm$^{-1}$ for $T \leq 100$ K is qualitatively similar to the data of the undoped BaFe$_2$As$_2$ [11] in shape as well as its temperature evolution, so that it is clearly the development of the SDW correlation with decreasing temperature below $\sim$ 100 K. Therefore, our sample simultaneously exhibits both the SC correlation as well as a short range SDW correlation in the widely separated frequency regimes.

For more convenient analysis, we convert $R(\omega, T)$ to the real part of conductivity $\sigma_1(\omega, T)$ using the Kramers-Kronig (KK) transformation [12]. For extrapolations of data, the high frequency part beyond 28,000 cm$^{-1}$ is extrapolated with the standard $1/\omega^4$ form. For the low frequency region below 15 cm$^{-1}$, the Hagen-Rubens (HR) formula is used in the normal state and $R(\omega, T) = 1$ is taken in the SC state. Figure 3 shows the results of the real part of optical conductivity, $\sigma_1(\omega, T)$, at various temperatures. The data of 200 and 300 K display a very broad Drude form centered at $\omega = 0$ and monotonically decrease until the interband transitions begin around 2000 cm$^{-1}$. Decreasing temperature to 100 and 50 K (still above $T_c$=36 K), three noticeable changes occur: (1) the spectral density around 700 cm$^{-1}$ builds up to form a broad hump structure; (2) concomitantly, the Drude peak rapidly sharpens ; (3) another new hump structure around 100 cm$^{-1}$, widely separated from the 700 cm$^{-1}$ structure, appears.

The first feature is the development of the short range SDW correlation. Instead of two distinct peaks resolved at 360 and 890 cm$^{-1}$, respectively, in the data of the undoped BaFe$_2$As$_2$ [11], a single broad peak centered around 700 cm$^{-1}$ is formed for $T < 100$ K and it became almost temperature independent from 50 K to lower temperatures. The second feature is the evolution of the formation of the coherent quasiparticles which interestingly coincides with the development of the SDW correlation in the mid-IR frequency range around 700 cm$^{-1}$. However, the most striking feature of our data is the third one, i.e., the appearance of the gap-like hump structure far above $T_c$ (50 and 100 K data), on top of the Drude spectra, in the range of 50 - 200 cm$^{-1}$, which is unmistakably a continuous evolution from the SC gap structure below $T_c$. This is the clear observation of a PG behavior solely caused by the SC correlation persisting up to temperatures of ~ $3T_c$. Below $T_c$, the data of 30, 20, 8, and 4 K show the opening of the SC energy gap with absorption edge at around 50 cm$^{-1}$ ($\Delta_S \approx 3.5$ meV) [13] and peaked around 100 cm$^{-1}$. This SC gap structure goes through a kind of dip around 200 cm$^{-1}$ and then is connected to the low end of the SDW correlation structure. This connecting dip region is actually the region where the large SC gap ($\Delta_L \approx 12$ meV) was identified in the optimally doped Ba$_{0.6}$K$_{0.4}$Fe$_2$As$_2$ [10]. Therefore, it may very well be that the large SC gap exists in our sample, too, but is masked by the low end part of the SDW structure around 200 cm$^{-1}$. Finally, the hump around 4,000 cm$^{-1}$ is due to the interband transitions and was also observed in the previous report [10, 11].

In order to further focus on the temperature evolution of the SC gap and related PG structures,

we subtract out the magnetic structure and the interband transition structure. The magnetic structure is weakly temperature dependent below 50 K and the interband transition is almost temperature independent. Therefore, we used the 4 K data for subtraction. Figure 4 shows the subtracted optical conductivity, $\sigma_{subt}(\omega, T)$, and the inset shows the Lorentz oscillators used to fit the subtracted structures. The subtracted optical conductivity below $T_c$, $\sigma_{subt}(\omega, T<T_c)$, exhibits the absorption edge around 50 cm$^{-1}$ and a peak at around 100 cm$^{-1}$ and then monotonically decreases afterward. Above $T_c$ up to 100 K, $\sigma_{subt}(\omega, T>T_c)$ shows the PG structure around 100 cm$^{-1}$ on top of the Drude spectra. The energy scale and the shape of this PG structure are self-revealing that it is a continuous evolution of the SC gap from below $T_c$. The peak shape above $T_c$ looks like it actually consists of two peaks of adjacent two SC gaps. This two peak structure becomes unresolved below $T_c$ because of their close adjacency [13, 14]. Now we consider a theoretical model that captures the temperature evolution of $\sigma_{subt}(\omega, T)$. $\sigma_{subt}(\omega, T)$ is the contribution from the free carriers which enter the SC condensate below $T_c$ and is released to the Drude spectra of conductivity above $T_c$. The optical conductivity below $T_c$ was first fitted by the standard Kubo formula with the SC gaps. And then above $T_c$, we adopted the phase incoherent preformed Cooper pair model [8] and followed the recipe of Franz and Millis [16] for the phase fluctuation average. The theoretical results are overlaid on the experimental data $\sigma_{subt}(\omega, T)$ in Fig.4. The theoretical conductivity $\sigma_{theory}(\omega, T>T_c)$ indeed captures the main features of the subtracted conductivity very well, namely, the PG structure as well as the Drude spectra.

In conclusion, we measured the optical reflectivity $R(\omega, T)$ of hole doped Ba$_{0.6+\delta}$K$_{0.4-\delta}$Fe$_2$As$_2$ ($T_c^{onset}$ =36 K) for temperatures from 4 K to 300 K and frequencies from 15 cm$^{-1}$ to 28000 cm$^{-1}$. We combined the THz and IR spectroscopies with a specially designed feed-back positioning system to maintain the overall uncertainty level less than 0.6 %. We identified the SC gap structure in 50 cm$^{-1}$ - 150 cm$^{-1}$ in the optical conductivity and its continuous evolution to the PG far above $T_c$, up to 100 K (~ 3$T_c$). Our data simultaneously displayed a development of the magnetic correlation structure in the mid-IR frequency range around 700 cm$^{-1}$ below 100 K. This coexistence and wide separation of the energy scales of the SC correlation and magnetic correlation structures led us to the conclusion that the PG structure in 50 cm$^{-1}$ - 150 cm$^{-1}$ is, beyond any doubt, caused by the SC correlation. Our observation, therefore, made Ba$_{0.6+\delta}$K$_{0.4-\delta}$Fe$_2$As$_2$ the first clear example of an unconventional high-$T_c$ superconductor exhibiting the PG behavior driven by SC correlation for the wide temperature range above $T_c$. Theoretical calculations based on the phase incoherent preformed Cooper pair model [8, 16] provided an excellent description of our optical conductivity data with the PG structure above $T_c$. Our findings put a constraint on the pairing mechanism of the iron-based superconductors and will shed new light on the origin of the PG in unconventional superconductors including the high-$T_c$ cuprates.

**Methods**

**Preformed Cooper Pairs model.** We consider an isotropic *s*-wave pairing state, and assumed that the SC condensate consists of two SC gaps $\Delta_{S1}$ and $\Delta_{S2}$ (two gap assumption is motivated from the shape of the PG structure above $T_c$ which looks like having two humps, but one gap fit of a similar quality is also possible with a substantially larger damping rate). The real part of the optical conductivity is calculated by a standard Kubo formula with sum of two bands (*a* = *S1, S2*) as follows,

$$\sigma_1(\omega, T) = -\frac{\sum_a \text{Im}\Lambda_{a,xx}(\omega)}{\omega} \quad (1)$$

with

$$\text{Im}\Lambda_{a,xx}(\omega) = \pi e^2 \int d^2k \, d\omega' v_{a,x}^2(k) \text{Tr}[\bar{A}_a(k, \omega + \omega')\bar{A}_a(k, \omega')] \times [f(\omega + \omega') - f(\omega)] \qquad (2)$$

where $v_{a,x}$ is the Fermi velocity along the *x*-direction of the electrons of the band $a = $ *S1, S2*, $f(\omega)$ is the Fermi-Dirac distribution function, and $\bar{A}_a(k, \omega)$ is the 2 × 2 spectral density matrix of Nambu green's function of the band *a* in the SC state defined as $\bar{A}_a(k, \omega) = -\text{Im}\bar{G}_a(k, \omega)/\pi$ and

$$\bar{G}_a(k, \omega) = \frac{\tilde{\omega}\tau_0 + \xi_a(k)\tau_3 + \Delta_a\tau_1}{\tilde{\omega}^2 - \xi_a^2(k) - \Delta_a^2} \qquad (3)$$

where the gap values $\Delta_a$ of the band "*a*" are quite straightforward to be determined without much ambiguity. We chose 50 cm$^{-1}$ for the smaller absorption edge $2\Delta_{S1}$ and 100 cm$^{-1}$ for the larger absorption edge $2\Delta_{S2}$. We need a phenomenological damping $\Gamma$ in $\tilde{\omega} = \omega + i\Gamma$ to fit the overall line shape and we used $\Gamma = 0.1\Delta_{S1}$ in our calculations. Without knowing $v_{a,x}$ and the DOS $N_a(0)$, the relative weight of the contribution from each band is taken as a fitting parameter. We chose the ratio of the spectral contributions of the *S1* and *S2* bands as 2 to 1.

Above $T_c$, we adopted the phase incoherent preformed Cooper pair model [8]. In order to simulate the phase fluctuations, we followed the recipe of Franz and Millis [16] and averaged the Nambu green's function with Doppler shift $\eta$ in the quasiparticle excitations as

$$\bar{\tilde{G}}(k, \omega) = \int d\eta \, P(\eta)\bar{G}(k, \omega - \eta) \qquad (4)$$

where the probability distribution of $\eta$ given by $P(\eta) = \sqrt{2\pi W} e^{-\eta^2/2W}$ with $W \simeq 3.48\alpha_v(T/T_C)\Delta^2$. $\alpha_v$ is a parameter derived from the XY-model and was estimated $\approx 0.009$ in the high-$T_c$ cuprates by Franz and Millis [16], for example. Here, we take the whole *W* as a fitting parameter and chose $W = 0.144\Delta_{S1}^2$ and $T = 1.2\Delta_{S1}$ which is about 2.1 $T_c^{BCS}$. The results of the theoretical calculations are plotted in Fig.4 in the main text.

**Acknowledgements** Y.S.K. was supported by the Nuclear R&D Programs funded by the Ministry of Science & Technology of Korea (2006-2002165 and 2009-0078025) and by Mid-career Researcher Program (NRF-R01-2008-000-20586-0) and Y.B. was supported by the Basic Science Research Program (NRF-2010-0009523). A.V.B. was supported by the US Department of Energy and by UCOP-TR01.

**Author contributions** Y.S.K designed and supervised the experiments. J.B.H, Y.R.J and B.H.M synthesized $Ba_{1-x}K_xFe_2As_2$ single crystal and J.B.H, Y.R.J, H.J.O, Y.Y.S, T.I, S.K carried out experiments and Y.B. and A.V.B. contributed to the theory and numerical calculations. Y.S.K., A.V.B. and Y.B. wrote the manuscript.

**Additional information** Supplementary information accompanies this paper on www.nature.com/nature. Reprints and permissions information is available online at http://npg.nature.com/reprintpermissions. Correspondence and requests for materials should be addressed to Y.S.K. (yskwon@skku.ac.kr) and Y.B. (ykbang@jnu.ac.kr).

**Figure 1. Temperature dependence of dc electrical resistivity of a $Ba_{0.6+\delta}K_{0.4-\delta}Fe_2As_2$ single**

**crystal.** Inset (a) is the enlarged dc electrical resistivity between 10 K - 60 K. Sharp superconducting transition is observed at $T_c^{onset}$ = 36 K. Inset (b) shows $(d^2\rho)/dT^2$ as a function of temperature. Inflection point is at 100 K.

**Figure 2. Reflectivity spectra, $R(\omega)$, at several temperatures between 4 K - 300 K up to 1,000 cm$^{-1}$.** At $T < T_c$, $R(\omega)$ approaches to unity in low frequencies due to the superconducting gap formation. Inset shows wide range spectra up to 28,000 cm$^{-1}$. Metallic behavior is observed for whole measured frequency range.

**Figure 3. Real part of optical conductivity, $\sigma_1(\omega, T)$, at several temperatures between 4 K - 300 K.** Dash-lines represent the Drude fitting for 300 K (grey dashes) and 50 K (dark grey dashes). At $T < T_c$, $\sigma_1(\omega, T)$ rapidly decreases and becomes gapped in low frequencies below around 50 cm$^{-1}$. Inset is the plot of plasma frequencies vs. temperatures $\omega_p(T)$ obtained from the Ferrell-Glover-Tinkham sum rule [15], $\omega_p^2 = \int_0^\infty \sigma_1(\omega, T)d\omega$. A cutoff frequency 2,500 cm$^{-1}$ is used instead of infinity to avoid the variation caused from the high frequency interband transitions. The $\omega_p^2$ is nearly constant in the PG temperature regions from 50 to 100 K and then it abruptly decreases below $T_c$ signaling the long range phase coherence of Cooper pairs as expected.

**Figure 4. Real part of optical conductivity, $\sigma_{subt}(\omega, T)$, after subtractions of the magnetic structure and the high frequency interband transition structure, for temperatures 4, 8, 20, 30, 50, and 100 K and theoretical conductivity $\sigma_{theory}(\omega, T)$ below (open squares) and above (open circles) $T_c$.** The theoretical calculations based on the preformed Cooper pair model are overlaid with the experimental $\sigma_{subt}(\omega, T)$. Two SC gaps are assumed as $2\Delta_{S1}$ = 50 cm$^{-1}$ and $2\Delta_{S2}$ = 100 cm$^{-1}$, respectively, and damping rate is chosen as $\Gamma = 0.1\Delta_{S1}$. Inset shows Lorentzian oscillators used for the subtracted spectra.

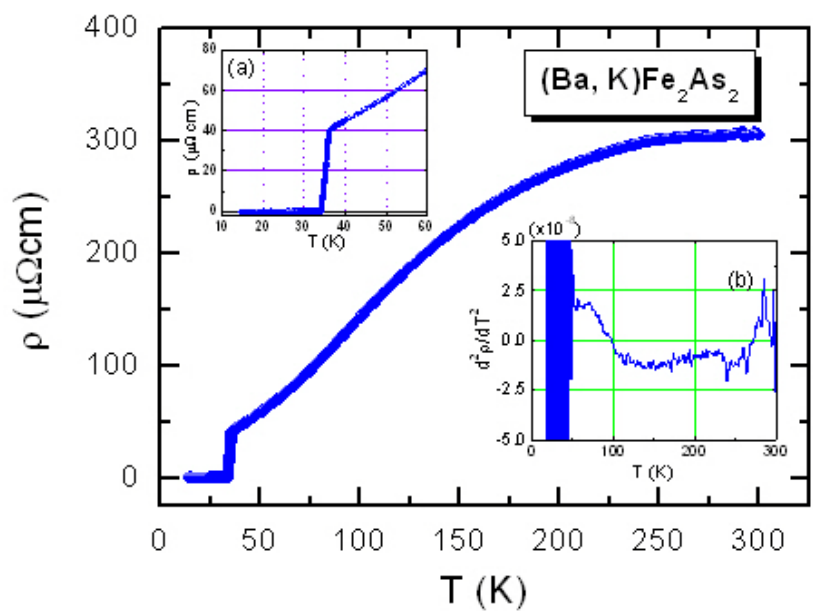

**Figure 1. Y.S. Kwon et al.**

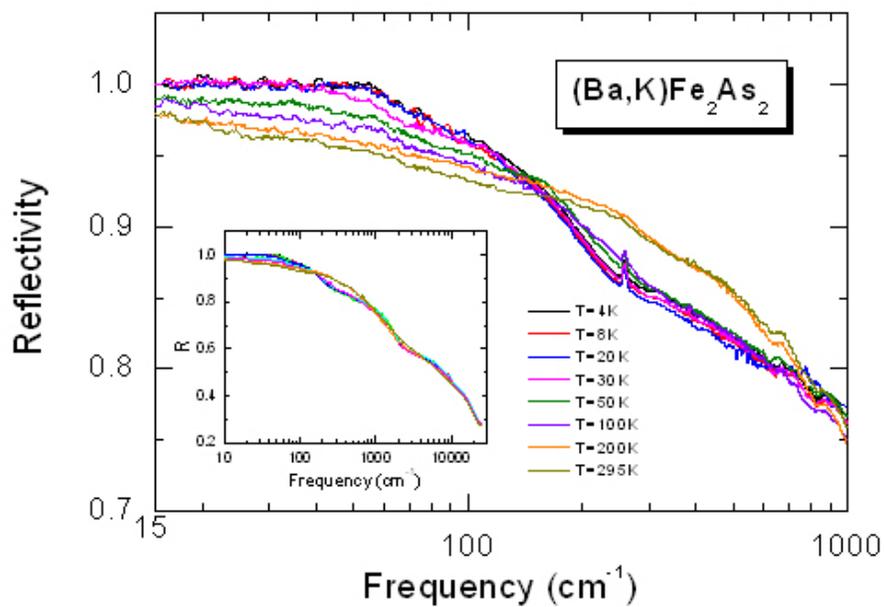

**Figure 2. Y.S. Kwon et al.**

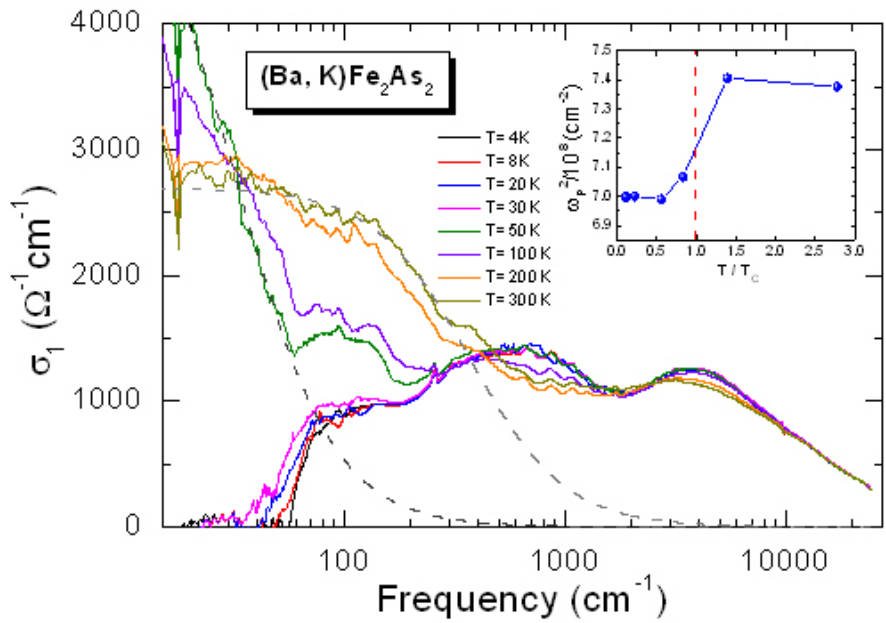

**Figure 3. Y.S. Kwon et al.**

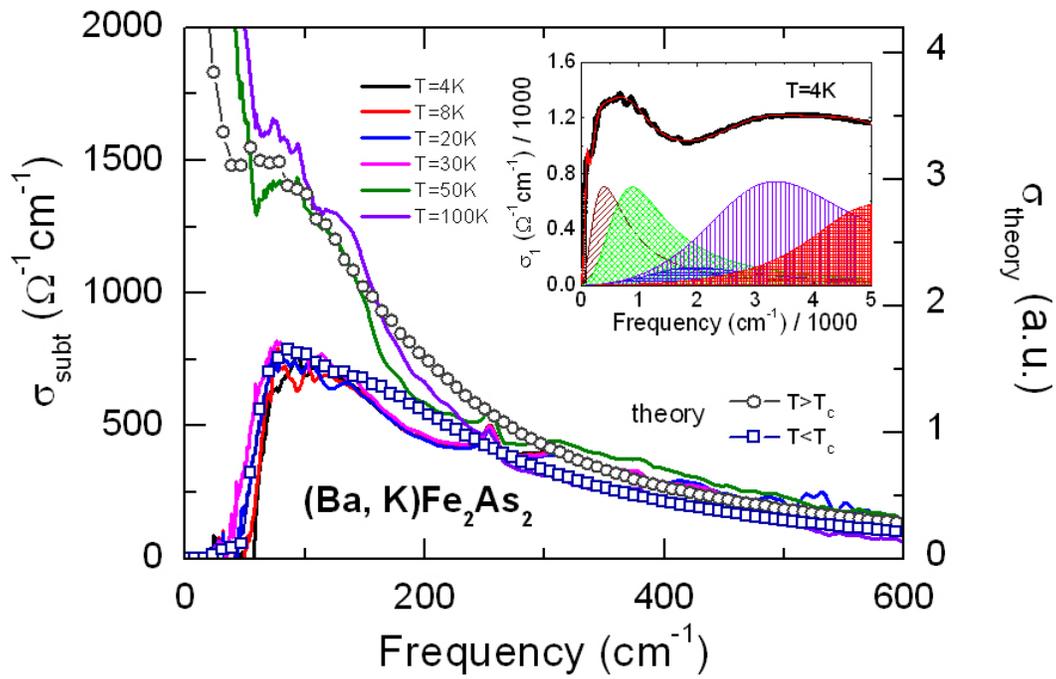

**Figure 4. Y.S. Kwon et al.**

**Supplementary Online Material**

**Experimental method to set up the absolute reflectivity and the feedback device to control the uncertainty.**

For the reference of the absolute reflectivity in THz and IR regions, an in-situ gold over-coating technique, which is a common method, was used in our experiment. In this setup, an Au-mirror as well as the sample was vertically mounted in the sample holder in cryostat. Moving the sample holder vertically, the intensity of the reflected beam from the sample and Au-mirror was sequentially measured using Si-bolometer cooled down to 4.2 K. After these measurements, Au was evaporated on the sample surface in-situ and the intensity of reflected beam was measured with the same method. The absolute reflectivity was calculated from the ratio of these two measurements.

In the reflectivity measurement of most pnictide superconductors including ours, the sample size is much smaller than the beam size (φ~8 mm). In this case, the interference of lights from the sample edge and the passing optical filters/windows is the main source of uncertainty of the measurement. In order to reduce this uncertainty, it is crucial to locate the same vertical positions of the targets (the sample, the Au-mirror and the Au-coated sample) through measurements. The usual method to find the vertical position by searching for the maximum intensity of reflection from the targets doesn't work because the intensity is nearly the same wherever the targets are located inside the beam area.

We resolved this problem by using the specially designed feedback method. As shown in Figure S1, we mounted a small piece of reference mirror in the opposite side of the sample and searched for the maximum laser intensity reflected from the mirror with the exactly same size of laser beam. This was performed by vertically moving the sample holder using a stepping motor with the resolution of 0.1 μm/step. Once the distance between the reference mirror and the sample and the distance between the sample and the Au-mirror were measured, we can find the same vertical position of the targets through our measurements and thus we could reduce the uncertainty level of 0.6 % and 0.3 % below and above 100 $cm^{-1}$, respectively.

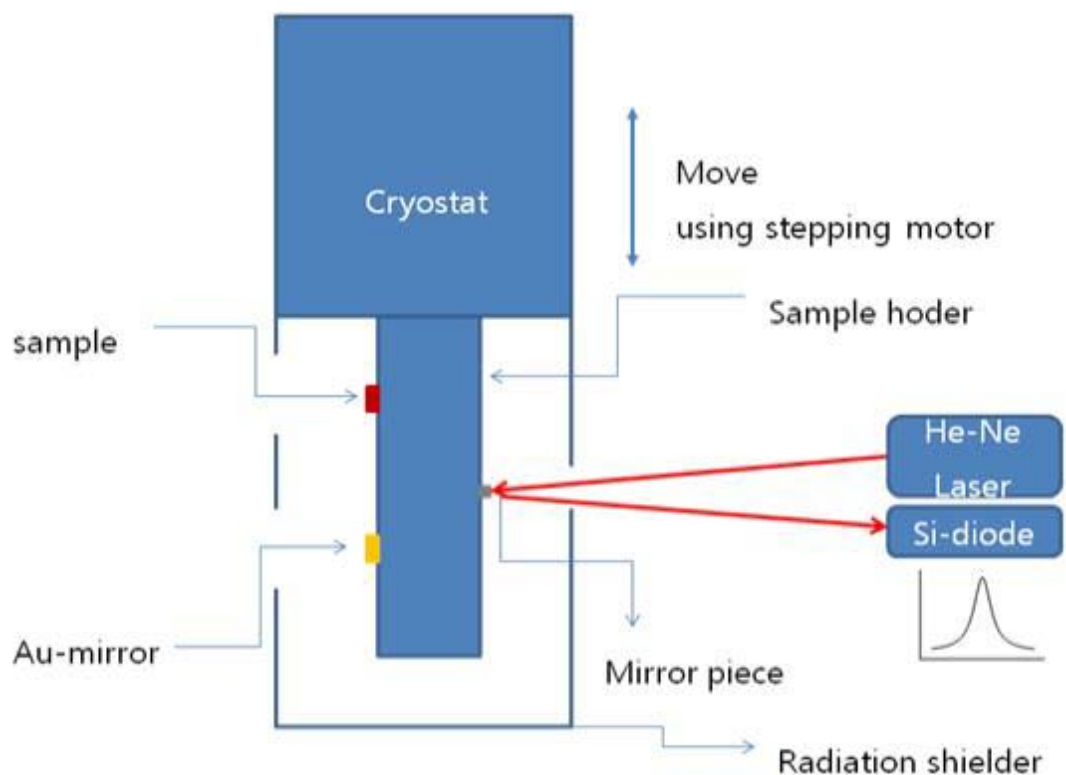

**Figure S1.  Diagram of measurement setup**